\documentclass{article}

\usepackage[preprint,nonatbib]{nips_2018}

\usepackage[utf8]{inputenc} 
\usepackage[T1]{fontenc}    
\usepackage{hyperref}       
\usepackage{url}            
\usepackage{booktabs}       
\usepackage{amsfonts}       
\usepackage{amsmath}
\usepackage{nicefrac}       
\usepackage{microtype}      

\usepackage{graphicx}               

\newcommand{\bincast}[1]{\mathrel{.}\mathrel{#1}}

\DeclareMathAlphabet\mathbfcal{OMS}{cmsy}{b}{n}

\title{Dynamic Automatic Differentiation of GPU Broadcast Kernels}

\author{
  Jarrett Revels\thanks{Corresponding author, jrevels@mit.edu} \\
  Massachusetts Institute of Technology \\
  \And
  Tim Besard \\
  Ghent University \\
  \And
  Valentin Churavy \\
  Massachusetts Institute of Technology \\
  \And
  Bjorn De~Sutter \\
  Ghent University \\
  \And
  Juan~Pablo Vielma \\
  Massachusetts Institute of Technology \\
}

\begin{document}

\maketitle

\begin{abstract}
We show how forward-mode automatic differentiation (AD) can be employed within larger reverse-mode computations to dynamically differentiate broadcast operations in a GPU-friendly manner. Our technique fully exploits the broadcast Jacobian's inherent sparsity structure, and unlike a pure reverse-mode approach, this ``mixed-mode'' approach does not require a backwards pass over the broadcasted operation's subgraph, obviating the need for several reverse-mode-specific programmability restrictions on user-authored broadcast operations. Most notably, this approach allows broadcast fusion in primal code despite the presence of data-dependent control flow. We discuss an experiment in which a Julia implementation of our technique outperformed pure reverse-mode TensorFlow and Julia implementations for differentiating through broadcast operations within an HM-LSTM cell update calculation.
\end{abstract}
\section{Introduction}
\label{sec:introduction}

In recent years, the prevalence of gradient-based optimization in machine learning (ML) has motivated an upsurge in the development of ML-specific modeling languages that incorporate automatic differentiation (AD) as a fundamental feature. However, contemporary ML research routinely seeks to utilize new modeling and optimization techniques that push these frameworks' AD capabilities to - and past - their limit. Both practical and exploratory implementations of such techniques demand advanced features such as nested differentiation, differentiation through data-dependent control flow, domain-specific hardware specialization, distributed parallelism, checkpointing, and more~\cite{maclaurin2015hyperparameter,chen2018tvm,lacey2016fpga,li2018adhalide,baydin2017adsurvey,paszke2017pytorch,abadi2016tensorflow}.

In the pursuit of solutions capable of incorporating such features, it has become clear that modeling languages' \textit{expressiveness} must necessarily be constrained for the sake of \textit{differentiability}. Recent endeavors~\cite{wei2017dlvm,breuleux2017myia,frostig2018jax,elliott2018categories,wei2018manifesto,innes2018zygote,pytorch2018scriptdocs,shaikhha2018diffprogarray} that explore this tradeoff have been strongly guided by well-established methods from programming language (PL) research, provoking the evolution of a new research area known as \textit{differentiable programming}. This is quite a natural development, as the narrative of traditional AD research has always been richly intertwined with PL and mathematical programming research.

Particularly relevant in this regard is the work of Siskind and Pearlmutter, whose stated vision aligns surprisingly well with the goals of contemporary differentiable programming research: ``Our vision: ...a unified intermediate language that supports both compiler optimizations and AD transformations for a variety of source and target languages.''~\cite{siskind2008vlad} When viewed through the lens of~\cite{naumann2000oja}, it can be concluded that ``optimal'' differentiation of such a language cannot be achieved via pure forward- or reverse-mode approaches, but rather demands a \textit{mixed-mode} approach. Achieving optimality, in this case, is often defined as minimizing the number of multiply-adds required to differentiate a given program via selecting the optimal mode for each subregion, and is known as the Optimal Jacobian Accumulation (OJA) problem. This problem has been shown to be NP-complete~\cite{naumann2008npcompleteoja}.

Despite this general theoretical intractability, mixed-mode AD offers a host of other advantages that can still be leveraged in practice by heuristically exploiting the local structure of the target language's primitive operations. This paper's primary concern is the application of this idea to a very common scientific computing primitive: the \texttt{broadcast} operation~\cite{vanderwalt2011numpyarray}. In \S\ref{sec:methodology}, we present the \texttt{broadcast} operation and a reverse-mode-interleavable forward-mode method for its differentiation that exploits the special sparsity structure of its total Jacobian. In \S\ref{sec:experiment}, we discuss an experiment that demonstrates our method's superiority over pure reverse-mode approaches for the differentiation of a data-dependent HM-LSTM cell update calculation on the GPU. \textbf{Via this paper, we wish to motivate the development of a new generation of differentiating compilers that do not operate purely in the forward or reverse modes, but rather choose the optimal mode for each target subprogram when such a choice is naively determinable from local structure.}

\section{Methodology}
\label{sec:methodology}

\subsection{Broadcast Operations}
\label{sec:methodology:broadcast}

\subsubsection{Broadcast Notation/Terminology}

Throughout this paper, we will append a period to a function to denote \textit{broadcasting} that function over its arguments. We define the broadcast of a function $b : \mathbb{R}^N \to \mathbb{R}^M$ as:
\begin{align*}
b.(\mathbf{X}_1 \dots \mathbf{X}_N) &= \texttt{broadcast}(b, \mathbf{X}_1 \dots \mathbf{X}_N) = \texttt{map}(b, \mathbfcal{X}_1 \dots \mathbfcal{X}_N) = \mathbf{Y}_1 \dots \mathbf{Y}_M
\end{align*}
Here, the $\mathbf{X}_j$ arguments are multidimensional arrays of arbitrary shape~\footnote{Note that scalars and single-element arrays are equivalent under this definition of \texttt{broadcast}.}, subject to the constraint that each dimension of any argument must either have the same length as that dimension in other arguments, or must have length $1$. Each $\mathbfcal{X}_j$ is equivalent to the corresponding $\mathbf{X}_j$, but where length-$1$ dimensions are ``copied'' along that dimension to match its maximum length across all $\mathbf{X}_j$, such that all $\mathbfcal{X}_j$ are of equal shape. The function $b$ is then mapped elementwise across all $\mathbfcal{X}_j$, resulting in the outputs $\mathbf{Y}_i$, each of which is the same shape as any $\mathbfcal{X}_j$.\footnote{Shrewd broadcast implementations
index directly into the $\mathbf{X}_j$ arguments to perform $b$ elementwise invocations (as is done in Eq.~\eqref{eq:bcastexample}) rather than explicitly materialize the $\mathbfcal{X}_j$ arguments.}

For example, broadcasting $b : \mathbb{R}^3 \to \mathbb{R}^2$ over an $n \times m$ matrix $\mathbf{A}$, a scalar $\alpha$, and an $n$-element vector $\mathbf{a}$ yields:
\begin{equation}
\begin{aligned}
b.(\mathbf{A}, \alpha, \mathbf{a}) =  
    &\left( \begin{bmatrix}
        b(A_{11}, \alpha, a_1)_1 & \dots  & b(A_{1m}, \alpha, a_1)_1 \\
        \vdots            & \ddots & \vdots                          \\ 
        b(A_{n1}, \alpha, a_n)_1 & \dots  & b(A_{nm}, \alpha, a_n)_1
    \end{bmatrix}\right.,  \\
    &\;\;\;\left.\begin{bmatrix}
        b(A_{11}, \alpha, a_1)_2 & \dots  & b(A_{1m}, \alpha, a_1)_2 \\
        \vdots            & \ddots & \vdots                          \\ 
        b(A_{n1}, \alpha, a_n)_2 & \dots  & b(A_{nm}, \alpha, a_n)_2
    \end{bmatrix}\right)
\end{aligned}\label{eq:bcastexample}
\end{equation}
To denote broadcast for binary infix operators, we prepend (instead of append) the period, e.g. $f.(\mathbf{X}_1) \bincast{+} g.(\mathbf{X}_2) = +.(f.(\mathbf{X}_1), g.(\mathbf{X}_2))$.

\subsubsection{Fusing Compositions of Broadcast Operations}

Assuming that a pair of broadcasted operations have compatible shapes and are relatively side-effect free, the broadcast of their composition generally obeys the following relation:
\begin{align}
g.(f.(\mathbf{X}_1 \dots \mathbf{X}_N)) &= (g \circ f).(\mathbf{X}_1 \dots \mathbf{X}_N) \label{eq:fusion}
\end{align}
In programs containing broadcast operations, Eq.~\eqref{eq:fusion} can be exploited to perform \textit{broadcast fusion}, a compiler-level optimization that transforms compositions of broadcast calls into a single broadcast call. This optimization imparts a couple of performance benefits. First, by obviating the need to compute and store intermediate results, broadcast fusion reduces memory usage, temporary allocations, and kernel invocations required to complete the computation. Second, broadcast fusion allows the fused broadcast operation to be parallelized without re-synchronization between intermediary broadcast operations~\cite{darte2000fusion,kennedy1994fusion}.

\subsection{Automatic Differentiation of Broadcast Operations}
\label{sec:methodology:diff}

\subsubsection{Multidimensional Dual Numbers}
\label{sec:methodology:diff:dual}

A common way to formulate forward-mode AD is via the algebra of \textit{dual numbers}. Dual numbers are similar to complex numbers, but instead of appending the imaginary unit $i$ to $\mathbb{R}$, the dual number algebra appends the infinitesimal perturbation $\epsilon$ where $\epsilon^2 = 0, \epsilon \neq 0$ to $\mathbb{R}$. Via Taylor series expansion, one can show that unary function application on dual numbers is defined as:
\begin{equation}
    f(x + y \epsilon) = f(x) + f^{\prime}(x) y \epsilon \text{ where } x,y \in \mathbb{R}
\end{equation}
This formulation is commonly implemented by defining a library of mathematical primitives that act on a \texttt{Dual} type consisting of the $(x, y)$ pair, e.g. \texttt{cos(Dual(x, y))} $\to$ \texttt{Dual(cos(x), -sin(x) * y)}. This \texttt{Dual} type can then be used to automatically differentiate any program composed of the defined primitives.

While this formulation is usually straightforward to implement, it is also quite limited - only a single scalar derivative can be calculated per call of the target function. To overcome this limitation, we can use an extended formulation of the dual numbers known as the \textit{multidimensional dual numbers}, which are defined as:
\begin{gather}
    x + \sum\nolimits_{i=1}^{k} y_i \epsilon_i \text{ where } \epsilon_{i}\epsilon_{j} = 0, \text{ s.t.} \\
    f\left(x + \sum\nolimits_{i=1}^{k} y_i \epsilon_i \right) = f(x) + f^{\prime}(x) \sum\nolimits_{i=1}^{k} y_i \epsilon_i
\end{gather}
Multidimensional dual numbers allow for a ``vector forward-mode'' implementation of gradient calculation, where orthogonal $\epsilon$ components are appended to orthogonal input components to compute their individual directional derivatives~\cite{pearlmutter2007lazy,revels2016forwarddiff}:
\begin{equation}
\mathbf{x} = 
\begin{bmatrix}
x_1 \\
\vdots \\
x_i \\
\vdots \\
x_k
\end{bmatrix} \to
\mathbf{x}_{\epsilon}  =
\begin{bmatrix}
x_1 + \epsilon_1 \\
\vdots \\
x_i + \epsilon_i \\
\vdots \\
x_k + \epsilon_k
\end{bmatrix}
\to
f(\mathbf{x}_{\epsilon}) = f(\mathbf{x}) + \sum_{i=1}^k \frac{\partial f(\mathbf{x})}{\partial x_i} \epsilon_i
\end{equation}
To extract the $\epsilon$ components as a tuple from a multidimensional dual number, we utilize the tangent extraction function $\mathbf{tg}$, defined as \begin{equation}
    \mathbf{tg_\alpha}(x + \sum_{i=1}^k y_i \alpha[\epsilon_i]) = (y_1 \dots y_k) \label{tgdef}
\end{equation}
Note that $\mathbf{tg}$ utilizes the notion of \textit{tagged perturbations}; $\mathbf{tg_\alpha}$ only extracts perturbations that are marked with the ``tag'' $\alpha$, represented here via the bracket syntax $\alpha[\epsilon]$. This tagging machinery is necessary (but unfortunately not sufficient) to avoid a class of AD bugs known as \textit{perturbation confusion}~\cite{siskind2008nesting,siskind2005perturbation,manzyuk2012confusion}.

\subsubsection{Sparse Forward-Mode Jacobians of Broadcasted Operations}
\label{sec:methodology:diff:broadcastjacobian}

Broadcasted operations generally take the form $b : \mathbb{R}^N \to \mathbb{R}^M$ where $N$ is the input arity and $M$ is the output arity. While $b$ might be broadcasted over millions of input elements, the arities $N$ and $M$ are generally relatively small (often $<10$). To automatically differentiate such a function in the forward-mode, we can define a Jacobian operator $\mathbf{D}$ using multidimensional dual numbers:
\begin{gather}
    \mathbf{D}(b) = (x_1 \dots x_N) \mapsto \mathbf{tg_\alpha}.(b(x_1 + \alpha[\epsilon_1], x_2 + \alpha[\epsilon_2], \dots x_N + \alpha[\epsilon_N])) \label{eq:doperatordef}
\end{gather}
where $\mathbf{tg_\alpha}$ is the tangent extraction function defined in Eq.~\eqref{tgdef}. Note that we are broadcasting 
$\mathbf{tg_\alpha}$ over the output tuple of $b$ in order to extract all components of the Jacobian. For example, for $b : \mathbb{R}^3 \to \mathbb{R}^2$, the definition expands to the following:
\begin{align*}
    \mathbf{D}(b) &= (x_1, x_2, x_3) \mapsto \mathbf{tg_\alpha}.(b(x_1 + \alpha[\epsilon_1], x_2 + \alpha[\epsilon_2], x_3 + \alpha[\epsilon_3])) \\
    &= (x_1, x_2, x_3) \mapsto \begin{pmatrix}
    \mathbf{tg_\alpha}(y_1 + \frac{\partial y_1}{\partial x_1}\alpha[\epsilon_1] + \frac{\partial y_1}{\partial x_2}\alpha[\epsilon_2] + \frac{\partial y_1}{\partial x_3}\alpha[\epsilon_3]) \\
    \mathbf{tg_\alpha}(y_2 + \frac{\partial y_2}{\partial x_1}\alpha[\epsilon_1] + \frac{\partial y_2}{\partial x_2}\alpha[\epsilon_2] + \frac{\partial y_2}{\partial x_3}\alpha[\epsilon_3])
    \end{pmatrix} \\
    &= (x_1, x_2, x_3) \mapsto \begin{pmatrix}
        \frac{\partial y_1}{\partial x_1} && \frac{\partial y_1}{\partial x_2} && \frac{\partial y_1}{\partial x_3} \\
        \frac{\partial y_2}{\partial x_1} && \frac{\partial y_2}{\partial x_2} && \frac{\partial y_2}{\partial x_3}
    \end{pmatrix}
\end{align*}
The following observation is frequently utilized throughout the rest of the paper:
\begin{gather}
    \text{Given } b : \mathbb{R}^N \to \mathbb{R}^M \text{ s.t. } \nonumber \\
    b.(\mathbf{X}_1 \dots \mathbf{X}_N) = \texttt{map}(b, \mathbfcal{X}_1 \dots \mathbfcal{X}_N) = (\mathbf{Y}_1 \dots \mathbf{Y}_M), \text{then } \nonumber \\
    \mathbf{D}(b).(\mathbf{X}_1 \dots \mathbf{X}_N) \to \left\{\texttt{diag}\left(\frac{\partial (\texttt{vec}(\mathbf{Y}_i))}{\partial (\texttt{vec}(\mathbfcal{X}_j))}\right) \mid i \in 1 \dots M, j \in 1 \dots N\right\} \label{eq:bcastjacobian} 
\end{gather}
where $\texttt{vec}$ is notation for vectorization (i.e. ``flattening'' the given tensor to a vector), and $\texttt{diag}$ is notation for extracting the diagonal of a square matrix. In other words, Eq.~\eqref{eq:bcastjacobian} directly computes all elementwise partial derivatives of the total Jacobian of $b.(\mathbf{X}_1 \dots \mathbf{X}_N)$. This approach exploits the sparsity structure imposed on the Jacobian by the broadcast operation, avoiding calculation of the zero-valued cross-element partial derivatives (the off-diagonal elements of the $\frac{\partial (\texttt{vec}(\mathbf{Y}_i))}{\partial (\texttt{vec}(\mathbfcal{X}_j))}$ matrices) by construction. Note, however, that if $b$ entails side-effects that induce cross-element dependence, then the total Jacobian is not fully recovered by this method, since the uncalculated cross-element partial derivatives may be nonzero in this case.

\subsubsection{Employing Forward-Mode Within Reverse-Mode}
\label{sec:methodology:diff:mixingmodes}

Eq.~\eqref{eq:bcastjacobian} has significant performance and programmability implications when exploited within larger reverse-mode AD computations, since it enables the differentiation of fully-fused broadcast subgraphs without requiring the construction of a backwards pass. Specifically, Eq.~\eqref{eq:bcastjacobian} can be employed to easily calculate and cache the intermediate Jacobian of the broadcast subgraph during the forward pass of the overall reverse-mode computation. This Jacobian can then be backpropagated during the reverse pass instead of backpropagating through the broadcast subgraph directly. In this way, Eq.~\eqref{eq:bcastjacobian} allows one to treat entire broadcast subgraphs as fused forward-mode primitives, obviating the need for reversible representations of these subgraphs.

\renewcommand{\arraystretch}{1.5}
\begin{table}
  \caption{Example Reverse-Mode Computation}
  \label{table:reversebcastexample}
  \centering
    \begin{tabular}{lll}
    Definition & Forward (Primal) & Reverse (Adjoint) \\
    \midrule
    $h(x, \mathbf{y}) = g(\mathbf{f}(x, \mathbf{y}))\quad\quad\quad$   &  &$\overline{w}_2 = 1 \text{ (seed)}$ \\
    $\mathbf{f}(x, \mathbf{y}) = b.(x, \mathbf{y})$ & $\mathbf{w}_1 = \mathbf{f}(x, \mathbf{y})\quad\quad\quad\quad$ & $\overline{\mathbf{w}}_1 = \overline{w}_2 \frac{\partial w_2}{\partial \mathbf{w}_1}$ \\
    $ b : \mathbb{R}^2 \to \mathbb{R}$ & $ w_2 = g(\mathbf{w}_1)$ & $\frac{\partial h}{\partial x} = \overline{\mathbf{w}}_1 \cdot \frac{\partial \mathbf{f}}{\partial x}$ \\
    $ g : \mathbb{R}^N \to \mathbb{R}$ & & $ \frac{\partial h}{\partial \mathbf{y}} = \overline{\mathbf{w}}_1 \bincast{\times} \frac{\partial f_i}{\partial y_i}$ \\
     $ x \in \mathbb{R}, \;\mathbf{y} \in \mathbb{R}^N$ & & \\
    \bottomrule
  \end{tabular}
\end{table}

To illustrate the use of Eq.~\eqref{eq:bcastjacobian} within a reverse-mode computation, consider the example defined in Table~\ref{table:reversebcastexample}. The left column defines the target function $h$, the center column expresses the primal forward pass of the computation, and the right column expresses the adjoint pass used to compute $\frac{\partial h}{\partial x}$ and $\frac{\partial h}{\partial \mathbf{y}}$. In the adjoint pass, the actual calculation of $\overline{\mathbf{w}}_1$ can be accomplished via the usual reverse-mode approach of decomposing $g(\mathbf{w}_1)$ into a reversible subgraph built from known primitive operations. The calculation of $\frac{\partial \mathbf{f}}{\partial x}$ and $\frac{\partial f_i}{\partial y_i}$, however, can be accomplished via Eq.~\eqref{eq:bcastjacobian} without requiring the construction of reverse-mode computation subgraph at all:
\begin{equation}
    \left(\frac{\partial \mathbf{f}}{\partial x}, \frac{\partial f_i}{\partial y_i}\right) = \mathbf{D}(b).(x, \mathbf{y}) \label{eq:computegderivs}
\end{equation}
Note that while it is mathematically useful to discuss $\mathbf{D}(b).(x, \mathbf{y})$ and $b.(x, \mathbf{y})$ as separate computations as we have done here, practical AD implementations leveraging this method can exploit the implicit computation of $b.(x, \mathbf{y})$ that occurs as part of computing $\mathbf{D}(b).(x, \mathbf{y})$. Instead of applying the $\mathbf{tg}_\alpha$ operator immediately as is done in Eq.~\eqref{eq:doperatordef}, the primal and dual computation results can be extracted simultaneously. In other words, the $\mathbf{w}_1$ step in forward pass code can be replaced with a step that simultaneously calculates $\mathbf{w}_1$, $\frac{\partial \mathbf{f}}{\partial x}$, and $\frac{\partial f_i}{\partial y_i}$ by simply invoking $\mathbf{f}$ with dual number inputs. 

Fusing the primal and derivative calculations in this manner avoids redundant computation, but requires that the resulting partial derivatives be cached until they are backpropagated in the reverse pass. This fusion, then, may not be desirable if there is not sufficient memory available to sustain such a cache. Conversely, computing $\mathbf{D}(b).(x, \mathbf{y})$ during the reverse pass redundantly computes $b.(x, \mathbf{y})$, but the resulting partial derivatives can be backpropagated immediately, and thus their storage can be freed (or reused) immediately. Ultimately, the choice of whether $\mathbf{D}$ should be applied in the forward pass or in the reverse pass depends on the memory/compute bandwidth of the overall computation.

\subsubsection{Forward-Mode vs. Reverse-Mode For $\mathbf{D}(b)$}
\label{sec:methodology:diff:forwardvsreverse}

The previous sections implemented $\mathbf{D}$ as a forward-mode differentiation operator, but $\mathbf{D}$ could have also been implemented via reverse-mode AD without invalidating Eq.~\eqref{eq:bcastjacobian}. Why, then, is forward-mode the better choice for this use case? The answer to this question can be summarized in three points:

{\bf 1:} If $N > M$, then reverse-mode is algorithmically superior to forward-mode. However, $b$ is generally low-arity, and in practice, forward-mode often outperforms reverse-mode for low-arity functions regardless of the $N/M$ ratio. There are two reasons for this. First, reverse-mode implementations often incur relatively high constant costs that are not amortized in the low-arity regime. Second, forward-mode's additional chain rule applications can be offset for low-arity functions by leveraging stack allocation schemes that make better use of cache bandwidth and allow for the exploitation of instruction-level parallelism~\cite{revels2016forwarddiff}.

{\bf 2:} If the target function contains data-dependent control flow, reverse-mode implementations must dynamically allocate the data-dependent regions of the computation graph\footnote{This requirement is not implementation-specific, but rather a hard theoretical limit; capturing intermediate values which depend on run time data will always require run time allocation in the general case, though certain optimizations may alleviate this burden in special cases. This requirement applies even to reverse-mode tools that claim to be ``tapeless'' by statically generating backwards pass code~\cite{merrinboer2017tangent,innes2018zygote}, or performing equivalent transformation via language-level constructs such as delimited continuations or closures~\cite{wang2018shiftreset}. As Pearlmutter and Siskind remark, it is ``impossible'' to ``eliminate the tape from reverse-mode AD'' because ``the tape stores intermediate values computed during the forward phase that are needed during the reverse phase.''~\cite{pearlmutter2008lambda}}. For low-arity functions, the overhead of dynamic trace allocation can easily dwarf the cost of the target function's primal evaluation. For broadcasted operations, this high overhead would be incurred for every elementwise invocation, rendering the reverse-mode approach in this case wholly unsuitable for the GPU where excessive dynamic allocation is infeasible.

{\bf 3:} Following from the previous point, using forward-mode for broadcast differentiation allows data-dependent control flow to occur within broadcasted scalar operations, thus avoiding several disadvantages inherent to vectorized control flow primitives currently employed by reverse-mode frameworks (e.g. TensorFlow's \texttt{where}~\cite{abadi2016tensorflow}). The first disadvantage is programmability; vectorized control flow primitives are often more cumbersome to use than their naive scalar counterparts. The second disadvantage is that many vectorized control flow primitives require computing untaken branches. While these primitives do have the benefit of clearly avoiding warp divergence on the GPU, the experiment described in \S\ref{sec:experiment} demonstrates that this benefit does not necessarily offset the cost of computing untaken branches on newer GPU architectures - especially if the difference in cost between branches is substantial - since newer architectures support executing different instructions across a warp without forcing serialized execution.

\section{Performance Experiments}
\label{sec:experiment}

In this section, we describe an experiment performed to compare this paper's forward-mode broadcast differentiation technique with existing reverse-mode approaches. Our test case for this experiment was a cell update calculation that occurs during the execution of a hierarchical multiscale LSTM (HM-LSTM)~\cite{chung2016hierarchical}, described in \S\ref{sec:experiment:hmlstm}. In \S\ref{sec:experiment:implementation}, we describe the three different AD implementations used to calculate gradients for our test case (TensorFlow-based reverse-mode, Julia-based reverse-mode, and Julia-based forward-mode). Finally, in \S\ref{sec:experiment:results}, we analyze GPU performance measurements obtained from benchmarking these implementations.

Note that all implementation, benchmark, and test code is publicly available in its  entirety at~\cite{paperrepo}.

\subsection{HM-LSTM Cell Update}
\label{sec:experiment:hmlstm}

The HM-LSTM cell update calculation is a real-world example of a broadcast operation that is amenable to differentiation via Eq.~\eqref{eq:bcastjacobian}. For a given time step $t$ and layer $\ell$, the update calculation for the cell $\mathbf{c}_t^\ell$ is:
\begin{equation}
    \mathbf{c}_t^\ell = \begin{cases} 
                            \sigma.(\mathbf{f}_t^\ell) \bincast{\times} \mathbf{c}_{t-1}^\ell \bincast{+} \sigma.(\mathbf{i}_t^\ell) \bincast{\times} \tanh.(\mathbf{g}_t^\ell) & \text{if }  z_{t-1}^\ell = 0, z_t^{\ell-1} = 1 \text{ (UPDATE)} \\
                            \mathbf{c}_{t-1}^\ell & \text{if } z_{t-1}^\ell = 0, z_t^{\ell-1} = 0 \text{ (COPY)} \\
                            \sigma.(\mathbf{i}_t^\ell) \bincast{\times} \tanh.(\mathbf{g}_t^\ell) & \text{if } z_t^{\ell-1} = 1 \text{ (FLUSH)}
                        \end{cases} \label{eq:cellupdate}
\end{equation}
where $\mathbf{f}$ and $\mathbf{i}$ are memory gates, $\mathbf{g}$ is a cell proposal vector, and $z$ is a boundary state.
  
We chose this operation as our experimental test case because it is self-contained, hinges on data-dependent control flow, has a substantial computational cost difference between branches, and is relevant to a machine learning audience. 

The benchmarks described in the following sections are primarily concerned with the calculation of $\frac{\partial \mathbf{c}_t^\ell}{\partial \mathbf{c}_{t-1}^\ell}$, $\frac{\partial \mathbf{c}_t^\ell}{\partial \mathbf{f}_t^\ell}$, $\frac{\partial \mathbf{c}_t^\ell}{\partial \mathbf{i}_t^\ell}$, and $\frac{\partial \mathbf{c}_t^\ell}{\partial \mathbf{g}_t^\ell}$.

\subsection{Tested AD Implementations}
\label{sec:experiment:implementation}

\subsubsection{Reverse-Mode TensorFlow Implementation}
\label{sec:experiment:implementation:tensorflow}

The first implementation tested in our experiment was a TensorFlow-based implementation derived from~\cite{finkelstein2017hmlstm}. This implementation makes use of TensorFlow's vectorized control flow primitive \texttt{where}, which eagerly computes both branches of the conditional statement before returning the branch specified by the given predicate. This primitive sidesteps actual branching and thus avoids two potential pitfalls discussed in \S\ref{sec:methodology:diff:forwardvsreverse}: dynamic trace allocation and warp divergence. As discussed in that section, avoiding these two perceived pitfalls comes at the cost of restricting the programming model and limiting opportunities for optimizations such as broadcast fusion.

A visualization of the implementation's post-optimization intermediate representation (IR) can be found in~\cite{paperrepo}, depicted as a computation graph in the High Level Optimizer (HLO) format. From this graph, it can be seen that TensorFlow's XLA compiler broke up the entire computation into six separate kernels, each representing a partially fused region of the forward and reverse passes, including broadcasted \texttt{select} operations that were generated from the initial code's \texttt{where} invocations.

\subsubsection{Reverse-Mode Julia Implementation}
\label{sec:experiment:implementation:reversejulia}

The second implementation tested in our experiment was a reverse-mode implementation in the Julia language~\cite{bezanson2017julia}. This implementation was directly derived from the HLO graph of the TensorFlow implementation described in the previous section. The intent was to exactly mirror TensorFlow's operations at the abstraction level of its HLO representation in order to better bridge comparisons between the reverse-mode TensorFlow and forward-mode Julia implementations. To accomplish this, the HLO graph operations were manually transcribed as native Julia code, additionally using the CUDAnative package to enable execution on the GPU~\cite{besard2017cudanative}.

\subsubsection{Forward-Mode Julia Implementation}
\label{sec:experiment:implementation:forwardjulia}

The third implementation tested in our experiment was a native Julia implementation of forward-mode broadcast differentiation as described by Eq.~\eqref{eq:bcastjacobian}. Like the reverse-mode Julia implementation, this implementation also employed the CUDAnative package to enable execution on the GPU. In many other ways, however, this forward-mode implementation differs substantially from the reverse-mode implementations.

The principle difference was the manner in which the primal calculation was expressed. While the reverse-mode implementations expressed control flow via vectorized primitives, the forward-mode approach allows the fusion of control flow into the broadcasted kernel without incurring the reverse-mode-specific performance penalties discussed in \S\ref{sec:methodology:diff:forwardvsreverse}. Thus, the forward-mode implementation's primal calculation was derived by using the relation from Eq.~\eqref{eq:fusion} to fuse Eq.~\eqref{eq:cellupdate} into a single broadcastable kernel:
\begin{align}
    \texttt{update}(c, f, i, g, z_1, z_2) &= \begin{cases} 
                            \sigma(f) \times c + \sigma(i) \times \tanh(g) & \text{if }  z_1 = 0, z_2 = 1 \text{ (UPDATE)} \\
                            c & \text{if } z_1 = 0, z_2 = 0 \text{ (COPY)} \\
                            \sigma(i) \times \tanh(g) & \text{if } z_2 = 1 \text{ (FLUSH)}
                        \end{cases} \label{eq:fusedupdate} \\ 
    \mathbf{c}_t^\ell &= \texttt{update}.(\mathbf{c}_{t-1}^\ell, \mathbf{f}_t^\ell, \mathbf{i}_t^\ell, \mathbf{g}_t^\ell, z_{t-1}^\ell, z_t^{\ell-1}) \label{eq:fusedcellupdate}
\end{align}
An implementation of the $\mathbf{D}$ operator (defined by Eq.~\eqref{eq:doperatordef}) was then applied to Eq.~\eqref{eq:fusedupdate} to calculate the required gradients. The $\mathbf{D}$ operator itself was implemented using the multidimensional dual number type provided by the ForwardDiff package, which represents a dual number as a pure Julia \texttt{struct} with two fields; one for the primal scalar, and one for a stack-allocated vector of perturbation coefficients~\cite{revels2016forwarddiff}.

\subsection{Experimental Setup}
\label{sec:experiment:setup}

Python code was executed using Python~3.6.3 with TensorFlow~1.5.0 and its XLA JIT compiler (which includes an unreleased version of LLVM close to LLVM~6.0). Julia code was executed using Julia~0.7.0-DEV.5025, built on LLVM 6.0. All required versions of all Julia packages used are publicly available: CUDAnative.jl version~0.8.1, CUDAdrv.jl~0.8.3, LLVM.jl~0.9.8, and ForwardDiff~0.7.5. We used the CUDA toolkit at version 9.1.85, in combination with NVIDIA driver 390.30 and Linux 4.13 from Ubuntu 17.10.

The implementations described in \S\ref{sec:experiment:implementation} were tested on NVIDIA Tesla V100, Tesla P100, and GTX 1080Ti GPUs in combination with 2 hexa-core Intel Xeon E5-2603~v4 CPUs and 64 GiB of DDR4 memory.

We measure the performance of individual implementations using the NVIDIA profiling tools from the CUDA toolkit. We only report kernel timings, excluding, e.g., memory transfers and CUDA API interactions, because the different implementations were designed to behave identically from an API point of view.

For the sake of accurate comparison, our Julia-based benchmarks followed TensorFlow's configuration where possible, e.g., using page-locked memory allocated asynchronously using the driver API, performing an identical amount of memory transfers, launch kernels identically (using at most 64 threads and a corresponding number of blocks), etc.

\subsection{Experiment Results and Analysis}
\label{sec:experiment:results}

\begin{figure}
  \centering
  \includegraphics[width=\linewidth]{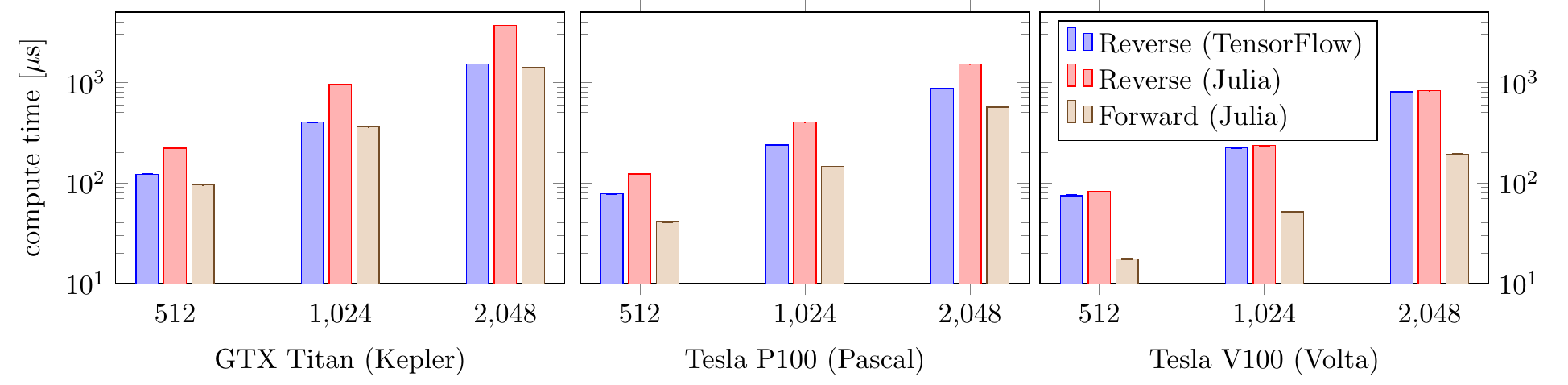}
  \caption{Total kernel compute times across different AD implementations}
  \label{img:perf:hmlstm}
\end{figure}

Fig.~\ref{img:perf:hmlstm} shows the execution times to compute the aforementioned derivatives for each implementation described in \S\ref{sec:experiment:implementation} across three generations of NVIDIA GPUs, with $\mathbf{c}_{t-1}^\ell$, $\mathbf{f}_t^\ell$, $\mathbf{i}_t^\ell$, and $\mathbf{g}_t^\ell$ taking $n \times n$ random 32-bit floating point matrix values and $z_{t-1}^\ell$ and $z_t^{\ell-1}$ taking $n$-element random 32-bit floating point vector values where $n \in \{512, 1024, 2048\}$. As can be seen in Fig.~\ref{img:perf:hmlstm}, the forward-mode Julia implementation features a speedup of $4.28$x, $2.66$x, and $2.60$x over the reverse-mode Julia implementation on the Volta, Pascal, and Kepler architectures, respectively. Compared to the reverse-mode TensorFlow implementation, these speedups are $4.18$x, $1.53$x and $1.07$x, respectively.

As mentioned in \S\ref{sec:methodology:diff:forwardvsreverse}, a substantial advantage of the forward-mode approach is that it avoids the computation of untaken branches by allowing data-dependent control flow to be fused within the broadcasted operation itself. However, this kind of fine-grained branching has traditionally been considered unfavorable for GPUs, which typically require threads within a so-called ``warp'' (a group of typically 32 threads) to execute in lockstep. If threads within a warp branch to different instructions, the hardware must execute both branches on all threads within the warp and mask out the results of untaken branches on each thread. This is known as \emph{warp divergence}, and can decrease performance significantly~\cite{bialas2015benchmarking}.

\begin{figure}
  \centering
  \includegraphics{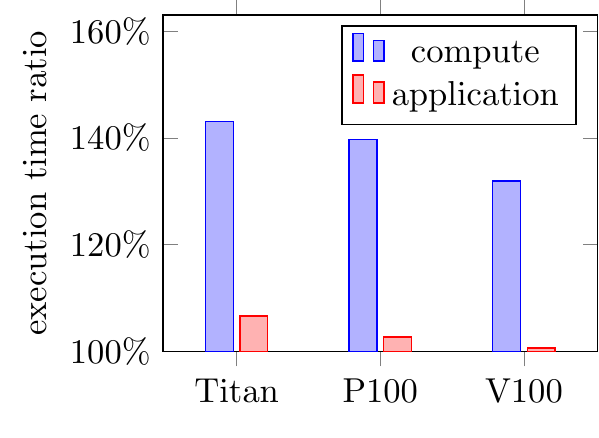}
  \caption{Forward-mode execution time with random control inputs vs. warp-uniform control inputs.}
  \label{img:perf:divergence}
\end{figure}

Fortunately, recent hardware improvements found on NVIDIA's Volta architecture can drastically mitigate the negative impact of warp divergence in many cases. This architecture enables independent thread scheduling by maintaining a program counter and call stack for every thread separately~\cite{nvidia2017v100}, thus allowing threads to execute different instructions without requiring serialized execution. The effects of this architectural improvement can be seen in Fig.~\ref{img:perf:divergence}, which shows the ratio of the forward-mode Julia implementation's execution time between uniformly distributed random control inputs and warp-uniform control inputs. Executing on a Tesla V100 GPU, the overhead of the thread-divergent implementation in terms of kernel execution time drops from $40\%$ to $30\%$ on Kepler and Pascal. When looking at total application execution time, this cost is even lower (below $1\%$), as kernels also execute faster on more recent hardware.

\subsubsection{Utilization Scaling With Increased Broadcast Arity}
\label{sec:experiment:results:arity}

\begin{figure}
  \centering
  \includegraphics{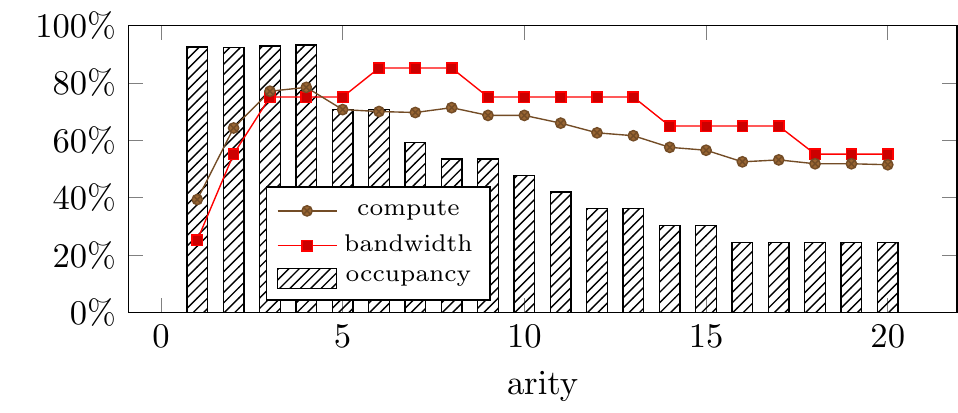}
  \caption{Effects of increasing operation arity on various utilization metrics on a Tesla V100 GPU.}
  \label{img:perf:arity}
\end{figure}

In addition to the main experiment, a different experiment was performed to explore how various indicators of GPU utilization scale as the arity of a forward-mode-differentiated broadcast operation increases.

As previously stated, the ForwardDiff package's multidimensional dual number implementation utilizes a stack-allocated vector to store perturbation coefficients. Recalling \S\ref{sec:methodology:diff:dual}, the length of every input, output, and intermediate dual number's perturbation vector is equal to the input arity of the target operation. Thus, increasing the input arity of the target operation increases register pressure. On CPUs, increased register usage can quickly result in excessive stack pressure, such that temporary values must be spilled into memory. On GPUs, however, many more registers are available; for example, the Tesla V100 contains 65,536 32-bit registers on each of its 84 Shared Multiprocessors (SM)~\cite{nvidia2017v100}. This advantage is offset by the large number of threads executing concurrently on the GPU, since each thread reserves a number of registers for exclusive use. The balance between active thread count and register usage is captured by the \textit{occupancy} metric, precisely defined as the ratio of active warps on an SM to the maximum number of active warps supported by the SM. With increased register usage, fewer warps can be allocated on each SM, and occupancy drops. 

To assess the impact of a broadcasted operation's input arity on the performance of its forward-mode differentiation by dual numbers, we designed an artificial benchmark and measured its achieved occupancy and effective hardware utilization: the computation of $\mathbf{D}(f).(\mathbf{x}_1 \dots \mathbf{x}_N)$ where 
\begin{align*}
    f(x_1 \dots x_N) &= \prod_i \tanh(g(x_i)) \\
    g(x) &= \begin{cases} 
        x & \text{if }x > \frac{1}{2} \\ 
        -x & \text{otherwise} 
        \end{cases}
\end{align*}
and each $\mathbf{x}_j$ is a $1024 \times 1024$ random matrix with 32-bit floating point elements. This benchmark provides a balanced workload for which it is easy to increase the arity $N$ and measure the subsequent effect on hardware utilization. Fig.~\ref{img:perf:arity} shows how occupancy drops steadily from 5 arguments on, at which point the amount of registers exceeds 32 and insufficient warps can be launched to satisfy the maximum number of concurrent warps per SM. Hardware utilization is initially limited by the low complexity of the kernel. It does not drop as strongly as the occupancy, since higher arity also increases the workload of the kernel, but at 18 arguments both compute and bandwidth utilization drop below 60\% and the kernel can be considered latency-bound due to low occupancy.

\section{Conclusion}
\label{sec:conclusion}

In this paper, we presented a reverse-mode-interleavable forward-mode method for the differentiation of \texttt{broadcast} that outperforms pure reverse-mode methods on the GPU and simultaneously obviates the need for reverse-mode-specific programmability restrictions on user-authored broadcasted operations. This mixed-mode technique is, in fact, already well-utilized in the Julia ecosystem. It was first introduced in 2016 by the ReverseDiff package (developed by this paper's first author)~\cite{revels2016reversediff}, whose original implementation of the method has since propagated to the Flux and Zygote packages ~\cite{innes2018flux,innes2018zygote}. This method was also recently employed for the differentiation of Julia broadcast operations on TPUs~\cite{fischer2018tpu}.

In the future, higher-order mixed-mode AD is likely to present interesting new challenges in the vein of perturbation/sensitivity confusion. For example, consider the forward-mode differentiation of the broadcast of a function that closes over variables naively tracked by a surrounding reverse-mode implementation. More research is needed to identify these potentially problematic scenarios and explore their ramifications.

Additional work has been planned to implement first-class mixed-mode AD for Julia within the upcoming Capstan package~\cite{revels2018capstan}, which will build on recently developed tools enabling third-party packages to extend Julia's compiler with new, context-specific behaviors by dynamically injecting code transformation passes into Julia's just-in-time (JIT) compilation cycle~\cite{revels2018cassette}.

\section{Acknowledgments}
\label{sec:acks}
Jarrett Revels was supported by MIT's NEC Corporation Fund for Research in Computers and Communications, MIT's Skoltech Next Generation Program and a Lawrence Livermore National Lab Subcontract. Tim Besard was supported by the Institute for the Promotion of Innovation by Science and Technology in Flanders (IWT), and by the Research Foundation -- Flanders (FWO). Valentin Churavy was supported by NSF DMS-1312831, Darpa XDATA, and an ARAMCO MITEI grant. The authors would also like to thank James Bradbury, Peter Ahrens, Mike Innes, Deniz Yuret, Ekin Akyürek and Simon Danisch for multiple helpful conversations and contributions to Julia's AD and GPU ecosystems.

\bibliography{references}
\bibliographystyle{plain}

\end{document}